\documentclass{article}

\usepackage{PRIMEarxiv}

\usepackage[utf8]{inputenc} 
\usepackage[T1]{fontenc}    
\usepackage{hyperref}       
\usepackage{url}            
\usepackage{booktabs}       
\usepackage{amsfonts}       
\usepackage{nicefrac}       
\usepackage{microtype}      
\usepackage{fancyhdr}       
\usepackage{graphicx}       
\graphicspath{{media/}}     

\title{Embodiment in multimodal large language models
\thanks{\textit{\underline{Citation}}: 
\textbf{Kadambi, A., Aziz-Zadeh, L., Damasio, A., Iacoboni, M., \& Narayanan, S. (2025). Embodiment in multimodal large language models (arXiv).}} 
}

\author{
  Kadambi A.$^{1,2,3}$\thanks{Correspondence to: \texttt{akadambi@ucla.edu}}, Aziz-Zadeh L.$^{1,2}$, Damasio A.$^{1}$, Iacoboni M.$^{3}$$^{\dagger}$, Narayanan S.$^{4}$$^{\dagger}$ \\
  $^{\dagger}$Dr. Narayanan and Dr. Iacoboni share co-senior authorship \\
  \\
  $^{1}$Brain and Creativity Institute, Dornsife College of Letters, Arts and Sciences, \\
  University of Southern California, Los Angeles, CA, USA \\
  $^{2}$USC Mrs. T.H. Chan Division of Occupational Science and Occupational Therapy, \\
  University of Southern California, Los Angeles, CA, USA \\
  $^{3}$Psychiatry and Biobehavioral Sciences, David Geffen School of Medicine, \\
  University of California, Los Angeles, CA, USA \\
  $^{4}$Google DeepMind, Zurich \\
}

\begin{document}
\maketitle

\begin{abstract}
Multimodal Large Language Models (MLLMs) have demonstrated extraordinary progress in bridging textual and visual inputs. However, MLLMs still face challenges in situated physical and social interactions in sensorally rich, multimodal and real-world settings where the embodied experience of the living organism is essential. We posit that next frontiers for MLLM development require incorporating both internal and external embodiment-- modeling not only external interactions with the world, but also internal states and drives. Here, we describe mechanisms of internal and external embodiment in humans and relate these to current advances in MLLMs in early stages of aligning to human representations. Our dual-embodied framework proposes to model interactions between these forms of embodiment in MLLMs to bridge the gap between multimodal data and world experience.
\end{abstract}

\keywords{Multimodal Large Language Models \and Embodiment \and Interoception \and Actions \and Alignment}

\section{Introduction}

As the next generation of large language models (LLMs) continues developing, Multimodal Large Language Models (MLLMs) or Large Multimodal Models, demonstrate impressive capabilities in interpreting text and images. These models are equipped with vast datasets that allow them to parse and generate complex multimodal representations. MLLMs achieve state-of-the-art performance on a range of multimodal tasks including image captioning and computer vision. MLLMs broadly include vision language models (VLM)s, including DeepMind’s Flamingo \cite{alayrac2022flamingo}, Gemini 2.5 Pro \cite{gemini2023family}, OpenAI’s GPT-5 \cite{openai2025gpt5}, GPT-4o \cite{openai2024gpt4o}, CLIP \cite{radford2021clip}, Amazon’s Nova \cite{amazon2024nova}, Meta’s Llama models \cite{touvron2023llama}, video language models such as VideoCLIP \cite{xu2021videoclip} and VideoLLaMa \cite{zhang2023videollama}, and many others. Due to their capabilities on a range of multimodal tasks, MLLMs have also synonymously been termed large world models or multimodal foundation models.

Still MLLMs face many challenges. Consider Moravec's paradox: when shown a slightly rotated image of a dot display easily recognizable as human (i.e., a point-light display) \cite{johansson1973visual}, the model states, "The image you sent appears to be a constellation." These models further perform poorly on action benchmarks \cite{cheng2025embodiedeval,yang2025embodiedbench,dang2025ecbench}, including for point-light displays \cite{kadambi2025evaluating}. Such shortcomings demonstrate clear difficulties with spatiotemporal and semantic grounding. With more training and better architectures, MLLMs are indeed likely to handle such limitations and improve performance on a range of multimodal tasks. However, MLLMs still lack any bodily experience. They interpret "heat" without ever feeling warmth, parse "hunger" without ever knowing need, and describe bodily actions without ever performing them or pantomiming them like humans do when they pantomime flying with their arms. Their understanding is statistical, lacking experiential grounding. Can a system truly comprehend the world if it has no body through which to experience it?

From birth, human bodies are felt from within and acted through in the world. Much of how humans learn about world dynamics is via embodiment, which positions the body as a fundamental building block of human cognition \cite{merleau2012phenomenology,varela1991embodied,damasio1996somatic}. While embodiment is often defined in terms of external interactions with the world—that cognition is situated and shaped by environmental context—it also includes the world inside the organism, or interoceptive experiences and other internal states \cite{merleau2012phenomenology,damasio2013nature, damasio2024homeostatic}. These internal signals, for instance hunger, temperature, and viscerality, are the basis for existence and survival \cite{damasio2013nature} and contribute to emotions \cite{damasio1996somatic}, memory \cite{quigley2021functions} and prosocial constructs such as emotional empathy \cite{carr2003neural,rizzolatti2005mirror,gallese1998mirror, iacoboni2009imitation}.

Language itself is to some extent grounded on such embodied experiences \cite{pulvermuller2005brain,clark2006language,gallese2008mirror,jirak2010grasping,feldman2004embodied}. In humans, language is not merely symbolic but shaped by experience. Through lived bodily experience, we form the internal worlds that support abstract reasoning and pragmatic understanding \cite{connell2018interoception}. Consider being asked, "Where is the salt?" at a dinner party. Understanding this utterance triggers a cascade of embodied representations—awareness of one's self location and bodily state, the hand affordances needed with salt-shakers, the action plan to pass the salt, the motivations and intentions of the speaker's indirect request, etc. This distinction is also apparent in how humans perceive objects and scenes. Seeing a cup can automatically and implicitly evoke imagery of drinking, seeing a lake can evoke swimming. Perception is inherently guided by these affordances and shaped by internal goals and bodily states. However, without internal and external models, none of these representations are triggered in MLLMs, which merely label and classify. For an LLM to perform the approximate behaviors would require it to have considerably better understanding grounded in internal and external models.

While there is no consensus on how to best conceptualize embodiment in artificial intelligence, we propose that embodiment can be usefully divided into these two complementary domains in cognitive science: (i) internal embodiment, or the internal dynamics (e.g., internal experiences, homeostatic control) that monitor internal states and maintain homeostasis and internal control and (ii) external embodiment, incorporating interactions with the external world (e.g., physical interactions). This paper argues that incorporating internal embodiment is not only timely but essential, especially as AI research has recently focused primarily on external embodiment via robotics and embodied agents. Integrating internal embodiment is required for models to move beyond token associations toward human-level generalization and prosocial alignment. Developing such internal models may be key toward more safer AI systems, while advancing machine intelligence with concrete proposals on mechanisms and preconditions for human embodiment.

\section{Internal states as states foundational to human life}

Most AI systems focus on external embodiment, such as action planning and physical execution, while lacking \textit{internal embodiment}, which we define as the persistent modeling of an agent’s own internal state. In biological systems, this includes behaviors such as interoceptive sensing and homeostatic regulation.

Internal bodily states are central to human life \cite{damasio2013nature}. Interoceptive information is in constant recurrent flow from internal organs to the nervous system. These states maintain an internal representation of the body or a ‘body schema’ \cite{merleau2012phenomenology,head1911sensory,graziano2002brain,gallagher1986body}.

Numerous internal feedback signals provide recurrent connections and endogenous control to maintain this body schema. Internal feedback loops are established in sensorimotor systems where the feedback originates from the system itself, rather than from the external environment \cite{desimone1995neural}. Such internal feedback is also useful for estimating the state of the system, controlling attention \cite{wolpert1995internal}, and maintaining homeostasis \cite{damasio1996somatic,sterling2012allostasis,friston2010free,deane2020dissolving}, where negative feedback loops keep the internal state stable within narrow ranges. Examples include thermoregulation, hormonal control, blood glucose regulation, and water balance.

The internal body model is also critical to social reasoning. Even when observing sparse visual actions such as point-light displays \cite{johansson1973visual}, humans effortlessly infer rich social attributes such as actions, intentions, and identities \cite{mather1994gender,kadambi2024individual,dittrich1996perception,manera2010inferring,kadambi2020understanding}, and even self-awareness \cite{kadambi2025self}. This ability emerges from a high degree of expertise in one’s own body, or motor experience.

Functionally, these internal states interact with multiple capabilities for action. For instance, humans routinely perform coordinate transformations. When we reach for a cup, we use an internal model to dynamically translate global (eye-based) coordinates to task-specific hand coordinates (like reach or preshape). When observing others, we similarly map their actions onto our own sensorimotor frame of reference. These transformations rely on predictive forward modeling, which anticipates sensory consequences before they occur \cite{wolpert1995internal}, partly because the delay in external feedback is prohibitive for real-time control. These recurrent and internal connections are ubiquitous in brains and serve multiple purposes, including attention, state monitoring, and maintenance for homeostatic control.

\section{Modeling the architectures}

\subsection{Neurobiology of embodiment in humans}

\begin{figure}
  \centering
  \includegraphics[width=0.8\textwidth]{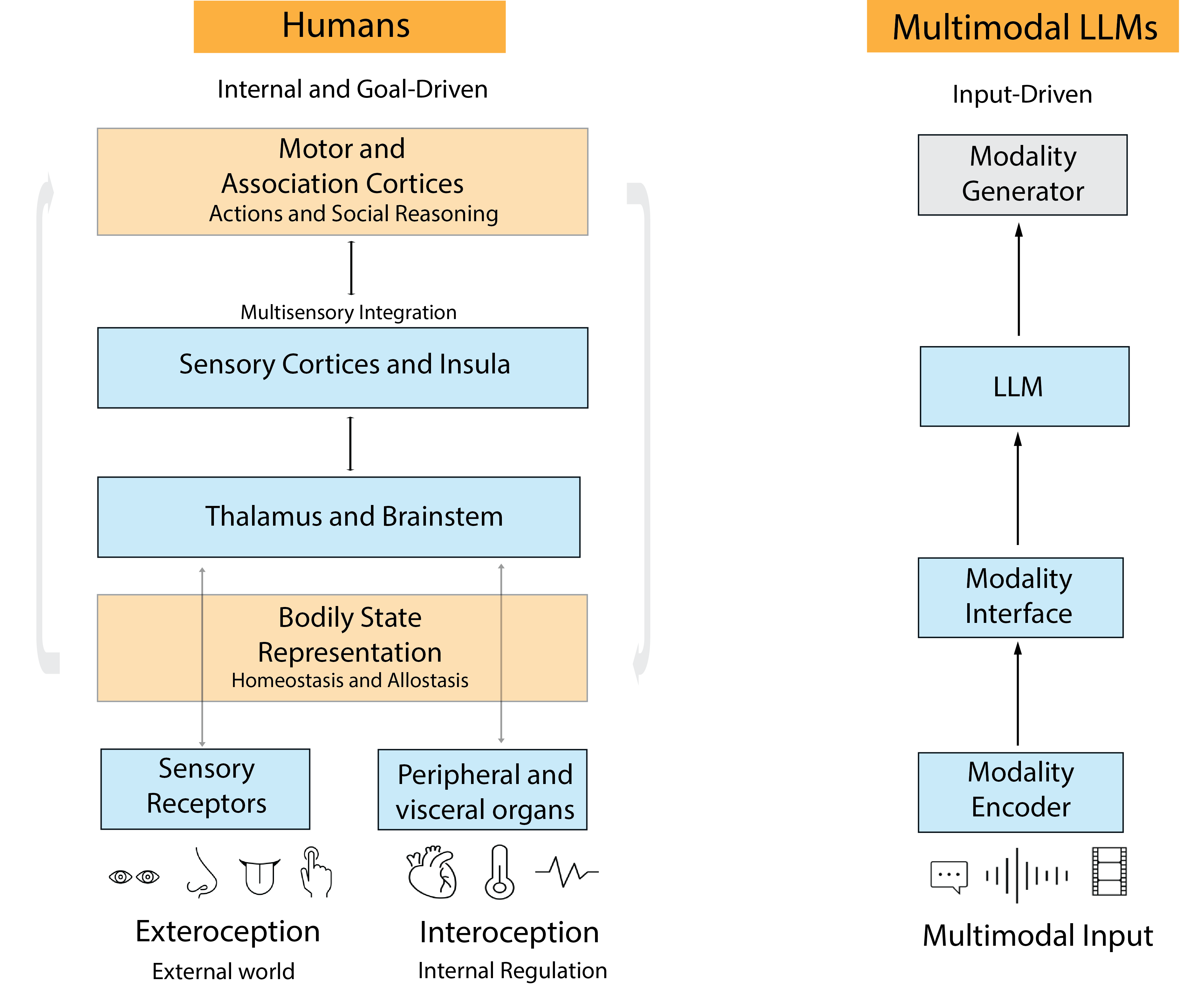}
  \caption{Standard embodied architecture in humans and multimodal LLMs. \textit{Left:} Neural architecture constructing embodiment in humans. Information flow is internal and goal-driven rather than input-driven. \textit{Right:} Standard implementation in MLLMs. Information flow is feedforward and based on the input stimulus.}
  \label{fig:architecture1}
\end{figure}

The internal capacities described above depend on well-defined brain systems. Beneath conscious access, signals from interoceptive sensors (e.g., chemoreceptors, mechanoreceptors) ascend via spinal and vagal pathways to brainstem and cortical hubs, including the insula and anterior cingulate. These structures generate integrated maps of bodily states (internal models) and via reciprocal loops, help maintain homeostasis \cite{berntson2021neural}. Signals from interoceptive areas also project to brain regions associated with motor functions, including the supplementary motor area and frontoparietal areas \cite{mufson1982insula,evrard2019organization,beckmann2009connectivity} to integrate internal bodily states with perception and actio\cite{craig2009feel,critchley2004neural} (see Figure~\ref{fig:architecture1}).

Internal models of our own bodies may also be used for cognition (i.e., understanding other people's actions, intentions, and experiences). For example, regions in the action observation network (AON; found in the inferior frontal gyrus, inferior parietal lobe, and precentral gyrus) engage not only when making actions but also when observing (or hearing) another individual performing a similar goal-directed action (i.e., mirror neurons, see \cite{rizzolatti2004mirror} for a review). This system may support social cognition (intention and action understanding) by mapping other people's actions onto our own internal motor representations, as a functional building block. Such shared circuits have also been found for other forms of social cognition, such as emotional empathy and vicarious touch \cite{aziz2018understanding}. These findings challenge earlier “sandwich models” of cognition, in which perception and action were conceptualized as independent modules connected only through an intermediary cognitive layer \cite{hurley1998consciousness,hurley2001perception}, and instead support the existence of an integrated social model of cognition.

There is also something fundamentally distinct about data arising from the body versus data arising from the external world. The former has a direct link to homeostasis and the maintenance of the organism’s integrity, imbuing it with an existential character. From a predictive processing perspective \cite{sterling2012allostasis,rao1999predictive}, interoceptive data must be predicted with higher prior precision, as maintaining life depends on keeping physiological variables within narrow ranges. In contrast, exteroceptive data, though critical for survival, can vary more widely and contextually, lending these modalities a more allostatic and anticipatory role \cite{CorcoranHohwy2018}. This is to highlight that interoceptive embodiment is not just an extra modality but a fundamentally different sort of data that must be actively controlled to support survival. 

Predictive processing frameworks have profound implications for language, which itself may be viewed as a hierarchical predictive system operating over generative internal models \cite{lupyan2015words}. Moreover, linguistic meaning may also depend on sensorimotor simulation: words like “grasp” or “run” engage neural circuits tied to their embodied referents \cite{pulvermuller2005brain,glenberg2012action,aziz2006congruent}. 

\subsection{Core Architecture of MLLMs}

MLLMs, however, lack the rich embodied and predictive representations seen in biological systems—their architecture instead reflecting classical “sandwich” models of cognition. This absence of interoceptive monitoring or explicit world modeling limits their ability to ground abstract linguistic concepts in physical experience \cite{aziz2006congruent}. Recall that even a simple utterance like “Where is the salt?” engages multiple layers of bodily inference in humans, such that concepts and syntactic structures emerge from embodied representations \cite{xu2025large,lakoff2025neural}. These layers of inference further influence higher cognition and the experience of prosocial constructs such as empathy and social reasoning \cite{man2019homeostasis,christov2023preventing}, which could be crucial for human alignment of AI systems.

The architecture of MLLMs traditionally integrates three main components: (1) \textbf{modality encoders}, (2) a \textbf{large language model}, and (3) a \textbf{modality interface} that connects the two. Modality encoders are specialized modules that transduce input data into machine-readable representations (typically vector embeddings) for the LLM. The LLM component provides the reasoning, memory, and symbolic manipulation capacities, while the modality interface bridges the encoder outputs with the LLM’s internal representations to enable multimodal communication. Some models additionally include a \textbf{generator module} that produces outputs in modalities beyond text (e.g., images, speech, or video).

These components are functionally separable. Modality encoders tokenize information by generating embeddings analogous to textual tokens or employ feature-level fusion to extract and integrate relevant signals across modalities. Feature-level fusion commonly leverages cross-attention mechanisms \cite{alayrac2022flamingo}, which align encoder outputs with internal LLM representations, or adapter modules that selectively emphasize task-relevant multimodal features. This modular organization enables flexible input–output mappings but, without internal feedback or interoceptive modeling, remains feedforward and externally driven.

\subsection{Constraints for embodiment in MLLMs}

Recent MLLMs demonstrate impressive capabilities such as optical character recognition-free reasoning, multimodal in-context learning, and dynamic inference-time computation. Moreover, post-training strategies like instruction tuning, reinforcement learning with human feedback and multimodal chain of thought\cite{wei2022chain} have significantly improved their reasoning ability and inference-time computation.

However, despite these advances, MLLMs still fall short of the perceptual richness and embodied inference characteristic of human cognition. While MLLMs excel at tasks such as image classification or object recognition (e.g., detecting edges, colors, and objects), these forms of feature extraction do not approximate the integrative nature of biological vision and action understanding. For instance, as discussed earlier, early MLLMs such as Gemini 2.0 failed to recognize simple point-light displays as human figures, even when the limb positions were exaggerated to make structure more salient. Later iterations, such as Gemini 2.5 Pro, show marked improvements, yet even minor spatial rotations of these displays can still disrupt recognition (see Figure~\ref{fig:architecture2}) and performance on dynamic point-light display benchmarks remains poor and well-below human performance \cite{kadambi2025evaluating}. To date, MLLMs also perform poorly on embodied benchmarks designed to test grounded reasoning, such as EmbodiedEval, EmbodiedBench, and ECBench \cite{cheng2025embodiedeval,yang2025embodiedbench,dang2025ecbench}. These evaluations measure aspects of embodied understanding, including spatial reasoning, affordance recognition, and environmental interaction. Improvements are on the clear horizon, though we raise the caveat that the evaluation benchmarks to date remain limited to external forms of embodiment, such as environmental model building and spatial reasoning.

\section{Toward Embodied MLLMs}

\begin{figure}
  \centering
  \includegraphics[width=0.8\textwidth]{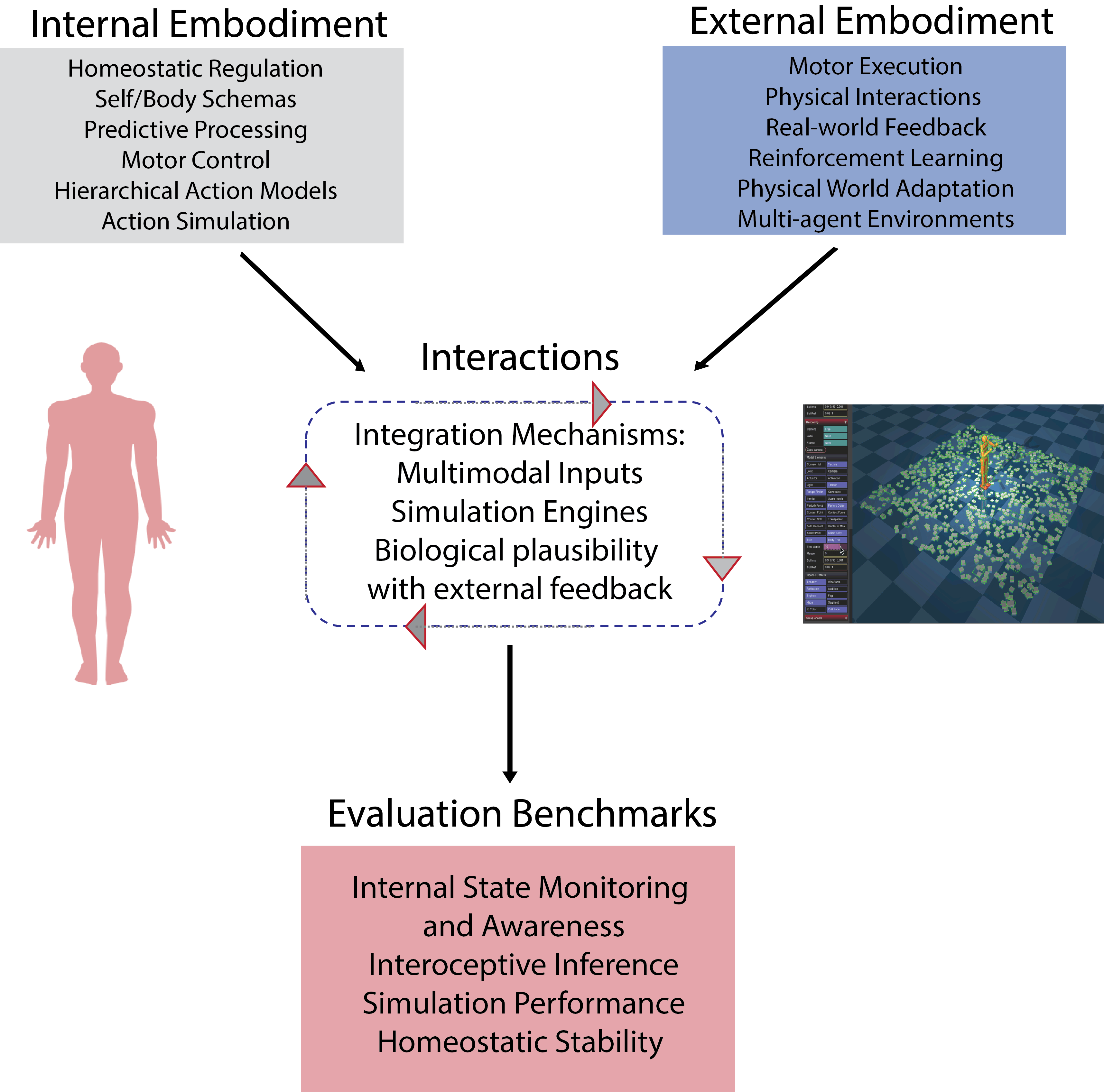}
  \caption{
  \textbf{Key Figure: Dual-embodiment framework.}
  A conceptual proposal for modeling interactions between \textit{internal} and \textit{external} embodiment in multimodal large language models (MLLMs). 
  The top left box illustrates possible empirical and computational approaches for modeling internal embodiment, such as homeostatic regulation, interoceptive feedback, and internal state prediction. 
  The top right box highlights strategies for modeling external embodiment, including embodied agents that interact physically or virtually with the world.
  The lower box outlines proposed benchmark paradigms for assessing dual embodiment, with emphasis on developing internal embodiment benchmarks (e.g., simulated homeostatic tasks, internal state regulation).
  The bottom right image depicts a frame retrieved from the \textit{Robot Report} (2023) showing DeepMind’s physics-based simulation environment MuJoCo \cite{todorov2012mujoco,deepmind2021mujoco}, an open-source engine for modeling physically grounded embodied agents.
  Related efforts such as the Genesis platform for embodied AI \cite{genesis2025} extend these principles by enabling large-scale multimodal simulations of sensorimotor interaction.
  }
  \label{fig:framework}
\end{figure}

To improve technical benchmarks on human-aligned MLLMs as well as to better inform the preconditions of embodiment in humans, we advocate for modeling both external and internal forms of embodiment in MLLMs (see Figure \ref{fig:framework} for an overview). First, we review current developments in these areas beginning with the external embodiment. While the distinction between external and internal has not yet been made in the literature, we do so here, noting that most AI research efforts primarily focus on the former. In the last section, we offer direct suggestions for their implementation.

\subsection{Current Directions: External Embodiment}

To our knowledge, all testing to date on embodiment in MLLMs has primarily focused on external embodiment, which aims to transform MLLMs from symbolic processors into systems capable of meaningful interactions in the world. External embodiment grounds MLLMs in environments that allow physical and contextual interactions. Embodied agents—whether in physical or virtual spaces—can perceive multimodal stimuli and physically engage in purposeful ways. Often referred to as "Agentic AI," these systems leverage human action-planning frameworks by incorporating goals and spatial awareness, and adapt well to environmental constraints\cite{durante2024agent,plaat2025agentic}. For instance, Action Space Adaptors (ASA)s transform standard MLLM outputs into measurable action outcomes \cite{szot2024grounding} such as robotic movements. Other MLLMs, such as EmbodiedGPT  \cite{mu2023embodiedgpt} directly learn policies by mapping sensory inputs (e.g., vision, proprioception) to actions, specifically for action planning and execution. These MLLMs can therefore execute sequentially complex tasks (e.g., opening cabinets) and perform meaningful world interactions. Other recent developments \cite{qiao2024agent} incorporate world knowledge generated from higher-level goals that are dynamically refined by the agent's local state.

For neuroscientists, these externally embodied MLLMs also provide a testbed to measure perception-action coupling to examine how embodied reasoning interacts with language, including how abstract reasoning may emerge from sensorimotor experience. They also allow researchers to test hypotheses about how sensorimotor grounding supports higher-level functions such as symbolic reasoning or goal inference. Across implementations, three core capabilities are prioritized: \textbf{(1)} perception, \textbf{(2)} planning, and \textbf{(3)} interaction with the external world \cite{durante2024interactive}.

\subsection{Future Directions: Modeling Internal Embodiment}

Internal embodiment in the biological sense has not yet been applied to MLLMs, yet they struggle with basic embodied tasks that appear external on the surface, such as human motion understanding (HMU-25; \cite{cheng2025embodiedeval,durante2024interactive}). 

As discussed, in humans, internal embodiment includes interoceptive processes of bodily states, as well as internal simulative and coupling processes that support empathy and prosocial behavior \cite{Singer2004} \cite{DecetyLamm2006}. Analogously, implementing internal state representations in artificial systems would allow MLLM-human interactions to function as coupled dynamical systems rather than as independent agents. In this framework, alignment and safety could emerge from shared sensitivity to homeostatic variables and their regulation, including empathy and emotional states \cite{friston2010free,man2019homeostasis}, which are of core importance for cooperative and prosocial behaviors.

Current MLLMs generally lack functional notions of internal control or internal state monitoring. Even Chain-of-Thought reasoning \cite{wei2022chain} and inference time compute approaches focus on sequential optimization rather than internal state dynamics. However, there have been interesting small-scale studies that appear to show benefits of neural network models predicting internal representations (i.e., their embeddings) as an additional task during standard classification tasks \cite{premakumar2024unexpected}. These metacognitive benefits to the models include efficiency through better weight regularization and improved performance. A related line of work with different objectives involves the use of hypernetworks \cite{chauhan2024brief}. While the goal is often rapid domain adaptation or distillation, not to self-monitor internal states, hypernetworks offer an intriguing possibility of constructing interventions and probes of internal states of the target network. In the following subsections, we review principles of internal embodiment in humans and assess their relevance and applications for MLLMs and cognitive science (see Table 1 for summary of implementations).

\subsubsection{Machinery for internal embodiment}

The machinery for internal embodiment is typically modeled in soft robotics, which aims to enhance bodily and biological plausibility in artificial systems (see \cite{man2019homeostasis} for an in-depth discussion; \cite{pfeifer2007self,lungarella2003developmental}). Recent developments in soft robotics have incorporated interoceptive hardware with artificial systems (non-LLMs). For instance, recent work \cite{yoshida2024emergence} demonstrates homeostatic internal systems capable of agentic behavior. These artificial agents might simulate intrinsic regulatory processes. Similar homeostatic-like processes are typically employed in other reinforcement learning agents \cite{yoshida2024empathic} and could be extended to MLLMs. 

But soft robotics also encapsulates cognitive machinery. For instance, self-maximizing principles like survival, and paradoxically vulnerability \cite{christov2023preventing} as well as prosocial goals, could improve functionality of these systems \cite{durante2024interactive,yoshida2024empathic}. With regard to human-alignment, recent developments in internal embodied AI systems demonstrate that prosocial behavior in reinforcement learning agents only occur when the homeostatic states of agents are coupled \cite{yoshida2024empathic}. Extending these mechanisms to MLLMs may not only enhance performance on embodied benchmarks (e.g., EmbodiedEval, EmbodiedBench) but also offer cognitive scientists a framework to model how internal regulation supports prosocial behavior.

\subsubsection{Recurrent organization}

Unlike recurrent neural architectures, MLLMs do not explicitly maintain or monitor internal states over time, other than the use of self-attention to attend to past and future tokens and activations within a temporal window \cite{vaswani2017attention}. Efforts to introduce recurrence, such as deep state-space models \cite{tiezzi2024state} are promising but early-stage. Additionally, sometimes the modality encoders or other modular components of pre-trained MLLMs are frozen during fine-tuning, preventing weight updating. This limits their ability to dynamically integrate information across modalities to adapt to new contexts, unlike humans \cite{alayrac2022flamingo,li2023fine}. Emerging frameworks like Deepmind’s adaptive agents \cite{adaptive2023human} aim to bridge this gap, but it remains an open-stage challenge to capture such functional flexibility as seen in biological systems.

\subsubsection{Internal modeling}

MLLMs also lack mechanisms for encoding and retrieving memories in structured, state-dependent ways. In humans, these memory processes are closely tied to internal models of bodily state. Interoceptive signals—such as those processed by the insula and anterior cingulate \cite{critchley2004neural}—give valence to memories and help shape how memories are formed and recalled \cite{quigley2021functions}. These internal embodied models support critical functions like attentional cueing, awareness, episodic replay, analogical reasoning, and autobiographical simulation \cite{gentsch2022clinical}. Embedding such interoceptive dynamics into MLLMs could enrich internal modeling by linking memory encoding to internal context.

Humans also simulate actions using internal models, such as forward models that predict sensory outcomes and comparator models that refine actions through feedback \cite{wolpert1995internal,seth2011opensim}. These mechanisms enable dynamic adaptation to changing environments. Internal models are also useful from a cognitive science perspective, since they can be used to generate and test various scenarios and their effects for action planning and control. While MLLMs for action anticipation also appear to learn structured action routines or skills, these learned representations remain limited in their flexibility. Their response is often directed by the prompt context or system instructions rather than goal-driven. 

Inference-time search and computation strategies (aka “thinking”) enable some degree of internal reasoning by exploring multiple action sequences to refine predictions during inference. However, currently these models work with fixed goals and often with verifiable answers, where actions are selected for a given context after being trained using reinforcement learning. Recent action anticipation models \cite{wang2025multimodal} combine sequence modeling and commonsense reasoning, based on learning statistical patterns and predicting future actions. While these predictive capabilities mark a step forward, they cannot adjust their predictions dynamically using real-time feedback. Internal state monitoring, runtime adaptation, recurrence, and internal feedback mechanisms are all crucial features of embodiment that are desirable but currently absent in MLLMs.

\section{Implementing interactions between internal and external embodiment}

We argue that the next frontiers in embodied MLLMs require moving from simple prompt/response such as in question answering to modeling complex tasks like action planning and execution in situated and dynamic environments. A prerequisite to this transition is the ability to also model internal states and their respective interactions. First, to embed internal states, proprioceptive sensors, such as motion capture sensors \cite{santina2019learning}, or pressure-sensors \cite{huang2020shape,best2016new} offer a partial bridge between internal and external domains. Interactions between internal and external embodiment could be incorporated in the form of proprioceptive feedback– robotic sensors (extending work by \cite{yoshida2024empathic}) —such as joint angle sensors or force sensors, to provide data about the robot’s internal state and update external actions. Recent implementations by DeepMind’s Gemini Robotics \cite{google2025gemini} include physical actions using robotic arms built on action models that are equipped with embodied reasoning abilities, such as spatial planning and reasoning, and pose and state estimation (Gemini Robotics-ER). Their internal software is also compatible with different embodied forms, including robotic (ALOHA-2) \cite{aldaco2024aloha,haddadin2022franka} and humanoid agents (Apptronik’s Apollo Robot; see \cite{noreils2024humanoid} for a review on humanoid robotics). While these approaches mark significant steps forward (e.g., 2x improvement in zero-shot task completion by using Robotics-ER over base Gemini 2.0), the evaluations remain limited to external embodied reasoning, such as the ERQA benchmark dataset \cite{gemini2025erqa}, leaving a major opportunity for internal embodied benchmarks (see Box 1).

Crucially, unlike proprioception, these interoceptive signals are not visible through external sensors and instead require internal modeling to maintain homeostasis. Explicit internal modeling of interoceptive variables—such as internal energy, thermal stress, and arousal—are important latent states within MLLMs or embodied agents. These states should dynamically change over time to influence reward landscapes, thereby improving alignment with human motivations and needs. Internal self-modeling will also become increasingly important in multi-agent environments, where distinguishing self from others either as an entity or a collective social group, and tracking internal states over time supports both autonomy and social coordination \cite{kim2013evolutionary}. As was mentioned earlier, even in fairly simple situations (such as image classification tasks), there are efficiency advantages in being able to predict internal representation as an ancillary task \cite{tiezzi2024state}. While the previous work stopped at showing weight regularization and performance gains for classification, it could be extended by modeling both internal and external interactions. That is, by using the constantly updated embedding predicted as the current state/context for action selection or other downstream tasks. To implement this, MLLMs could be trained on multisensory datasets including proprioceptive, environmental, and simulated interoceptive inputs. Data from motion capture systems, robotics, or simulated engines like Genesis \cite{genesis2025} or embedded on platforms like Unity (https://unity.com/) could provide the physical grounding necessary for such integration.

Embodied actions also deal with reasoning and control at multiple time scales. Humans actively simulate and infer higher-level goals and intentions through several sequences of actions. For instance, grasping an object is not only a motor action but also part of a larger intention, with the aim to ‘drink from a cup’ \cite{grafton2007evidence,iacoboni2005grasping}. Designing MLLMs to effectively replicate this hierarchical processing would enable more complex action inference and action simulation (see Proc4Gem; \cite{lin2025proc4gem}). A standard approach is to use a layered control architecture \cite{matni2024quantitative} where there is a hierarchical decomposition of actions from continuous time feedback controllers and reflexes all the way to longer-term and higher-level goals and intents. 

Promising steps in these directions include internal multimodal chain-of-thought approaches \cite{wei2022chain}, which demonstrate how models can simulate future world states across modalities to support reasoning and action planning. We would like to propose, in addition to these exciting upcoming developments, the need for special attentional and internal feedback (from higher latent layers to input layers) mechanisms to represent, monitor, predict, and adjust relevant internal state variables. This would likely require modifications to the architecture (including recurrence) as well as to the reinforcement learning algorithms we use to incorporate state values, actions, and rewards, as well as additional data from simulated or actual interactions. This has been explored successfully for games with clear, verifiable rewards by systems such as AlphaZero and MuZero \cite{schrittwieser2020mastering} where reinforcement learning from self-play experience proved superior to human data-based learning. The agents produced in this manner displayed superior transfer learning capability in novel game domains. Scaling these ends to produce better embodied benchmarks (with complex internal states and actions) in real-world robotic domains is still an open problem requiring further research. These advances may also yield not only more flexible and adaptive AIs, but also computational analogs for key cognitive functions long-theorized but difficult to isolate in biological systems.

For cognitive scientists, these developments offer key testbeds to examine long-standing theories of embodied cognition \cite{friston2010free}. For example, predictive processing models can be evaluated by probing how embodied MLLMs resolve mismatches between expected and actual feedback. Comparator models of motor control can be instantiated through simulated proprioceptive error correction. Multi-agent MLLMs may even offer ways to test the emergence of joint attention or perspective-taking. These frameworks can ultimately help clarify how internal and external dynamics are integrated to support adaptive and agentic behaviors.

\begin{table}
\caption{Summary of modeled embodiment in MLLMs}
\centering
\begin{tabular}{p{2.5cm}p{2.5cm}p{3cm}p{3.5cm}p{3cm}}
\toprule
\textbf{Feature} & \textbf{Humans} & \textbf{Current MLLMs} & \textbf{Proposed Improvements} & \textbf{Example Implementation} \\
\midrule
\textbf{Internal Embodiment} & Rich sensorimotor experiences, interoception, predictive models & No built-in body model, lacks adaptation and internal feedback & Integrate hierarchical predictive models with homeostatic feedback & Homeostatic coupling to create AI with intrinsic motivation \\
\midrule
\textbf{External Embodiment} & Active sensorimotor interaction with physical world & Generally lack physical interaction, relies on datasets & Implement embodied learning and real-world sensory feedback loops & Train AI in virtual and real-world environments \\
\midrule
\textbf{Interactions} & Humans continuously refine actions based on real-time feedback & MLLMs use static datasets and are largely pre-trained & Develop closed-loop learning architectures & Simulation engines, multimodal datasets, internal action hierarchies \\
\bottomrule
\end{tabular}
\label{tab:embodiment}
\end{table}

\section{Concluding Remarks}

Developing safer, human-aligned world models requires modeling internal states alongside external embodiment. Models capable of simulating and understanding embodiment in its multiple aspects can better navigate their environments and may reduce the risk of misaligned or maladaptive behaviors, including those that appear sociopathic or harmfully indifferent to others \cite{christov2023preventing}, and increase cooperative behaviors \cite{christovmoore2025contingencies}. This article draws from principles of internal and external embodiment in humans as a potential means of resolving embodied limitations in MLLMs (see outstanding questions) to promote trust, greater alignment, and safety between AI systems and human users.

\section*{Human-aligned internal benchmarks}

Most MLLM benchmarks focus on external tasks like object manipulation or spatial reasoning. To advance toward human-aligned and self-regulating AI, we propose several benchmarks that evaluate internal embodiment.

1. \textbf{Simulated Homeostatic Tasks} \\
MLLMs could be tested in environments with internal state variables (e.g., energy, temperature) as well as stressors (e.g., pain) that fluctuate over time. Agents must maintain stability by choosing actions like “resting” or seeking “food,” even when they conflict with external task goals. Performance would reflect the ability to prioritize internal stability and trade-offs.

2. \textbf{Prosocial Benchmarks} \\
MLLMs could be placed in multi-agent environments where one’s internal state affects the other’s. Helping behaviors or resource sharing in response to distress would reflect prosocial inferences driven by internal modeling \cite{seth2025conscious}. Empathy benchmarks could ultimately test whether MLLMs simulate what another agent might be “feeling.”

3. \textbf{Self-Monitoring and awareness} \\
Two questions arise around internal modeling of awareness. One is the ability of MLLM to consistently simulate specific personas, roles, and styles when instructed. For instance, prompts such as “Write my biography as if you were a Victorian poet,” or more recently “Draw an image of me in the Ghibli style,” allow the MLLM to consistently respond and generate text or images in the requested style. These role-playing cases are extremely interesting aspects of MLLMs, though such steering or style transfer can arise without any internal state estimation. It was commonly known that when asked to identify itself, DeepSeek (and other MLLMs) would commonly reply “ChatGPT,” choosing the most probable token sequence in the context \cite{lunden2024why}. The second question concerns self-critique. This is an extremely useful capability in MLLMs that has been exploited in using argumentation as a way to arrive at correct answers for a given prompt. Yet critiquing a previous response does not require internal state estimates or models since the input is the observable output of a previous prompt/context. There is some evidence that LLMs exhibit a self-preference bias \cite{panickssery2024llm,wataoka2024self} in judging their output as better than other model outputs, possibly because of stylistic similarity or a bias for reduced perplexity in responses \cite{wataoka2024self}, but fail to transfer to many self-recognition tasks \cite{davidson2024self}. Recent work such as SelfIE (Self-Interpretation of Embeddings) \cite{chen2024selfie} uses a framework that enables LLMs to interpret their own embeddings in natural language by passing the embedding as context and asking specific queries, offering a possible approach to self-monitoring.

\section*{Acknowledgments}

We thank Caitlin Nguyen, Nicole Caballa, and Elisa Liu for assistance in organization of the references and manuscript feedback. We thank Google DeepMind for intellectual and financial support.

\section*{Declarations of Interest}

Srini Narayanan is an employee of Google DeepMind. The authors have no other competing interests to declare.

\section*{Glossary}

\paragraph{Active Inference:} An extension of predictive processing grounded in the free energy principle. Active inference proposes that agents not only update internal beliefs to minimize prediction error but also take embodied actions to actively change their environment to better match those predictions.

\paragraph{Body Schema:} Simplified representation that lacks, for instance, a representation of the precise body dynamics (torques, forces, stress) on joints muscles. It corresponds to an implicit knowledge of the location and movement of our limbs along with a sense of a fluid boundary that separates self from others.

\paragraph{Embodied Reasoning Question Answering (ERQA):} Benchmark that focuses specifically on embodied capabilities using interactions with the physical world.

\paragraph{Forward Model:} For both our own and other people's actions, the forward model is essential to smooth correct movements/actions because it provides a prediction of the movement, which is compared with incoming sensory and proprioceptive feedback to generate error signals that guide correction.

\paragraph{Hypernetworks:} Neural networks that generate weights for another neural network, called the target network. They act as a meta-learner, producing weights that can be dynamically adjusted based on different tasks or inputs. They offer potential for faster training and continual learning and adaptation.

\paragraph{Inference-time Compute:} Often referred to as "thinking" in AI contexts, refers to the computation a model performs after it has been trained to generate outputs in response to new inputs.

\paragraph{Instruction Tuning:} Method for improving large language models by fine-tuning them on datasets of natural language instructions that are paired with task-specific responses.

\paragraph{Interoception:} Nervous system's process of sensing and integrating internal bodily signals, which actively shapes awareness of the body's internal state. Emerging early and pre-reflectively, it is the foundation for embodiment and consciousness.

\paragraph{Moravec's paradox:} AI's difficulty in often performing tasks that humans find easy, while excelling at tasks that humans find complex. The in-text example shows how the ability to recognize a point-light display from a static frame when the limbs are positioned away from the body is clear-cut and trivial for humans.

\paragraph{Multimodal Chain of Thought:} Step-by-step reasoning in AI that integrates information across modalities (e.g., vision, language, touch).

\paragraph{Multimodal Large Language Models (MLLMs):} Or Large Multimodal Models, the multimodal and next generation of large language models, demonstrate impressive capabilities on a range of multimodal stimuli, such as computer vision, video and image captioning and generation. Due to their breadth, MLLMs have synonymously been termed large world models or multimodal foundation models.

\paragraph{Optical Character Recognition-free Reasoning:} Ability of an MLLM to understand and reason about textual content embedded in images without using a separate optical character recognition step. Instead, the model processes the visual and textual elements jointly.

\paragraph{Point-light Display:} Typically around a dozen dots depicted on key joints of the human body. Humans and even newborns are able to easily and immediately infer human body structure from the sparse visual cues.

\paragraph{Predictive Processing:} Theoretical framework in cognitive neuroscience that positions the brain as a hierarchical prediction machine. In this view, the brain continuously generates top-down predictions about incoming sensory input, which are then compared with actual input. The brain's ultimate goal is to minimize this discrepancy, or prediction error.

\paragraph{Recurrence:} Network architecture where past outputs are fed back into the model, which enables the maintenance of internal states over time.

\paragraph{Reinforcement Learning with Human Feedback:} Training technique that fine-tunes language models based on human preference signals. Human annotators compare model responses, and these preferences are used to train a reward model. The language model is then updated using reinforcement learning to generate more human-aligned outputs.

\paragraph{Self-attention:} Key mechanism in large language models that allows a model to weight the relevance of different tokens in a sequence relative to each other. Each token (e.g., a word) is compared with all others to compute attention scores. This allows the model to decide which parts of the input have the highest attention scores and therefore the most relevant for deriving meaning in context.

\paragraph{Soft Robotics:} Subfield of robotics that aims to create artificial intelligence from flexible, deformable materials inspired by biological systems and principles.

\clearpage                      
\section*{Supplementary Figures}

\setcounter{figure}{0}
\renewcommand{\thefigure}{S\arabic{figure}}

\begin{figure}[!htbp]           
  \centering
  \includegraphics[width=0.8\textwidth]{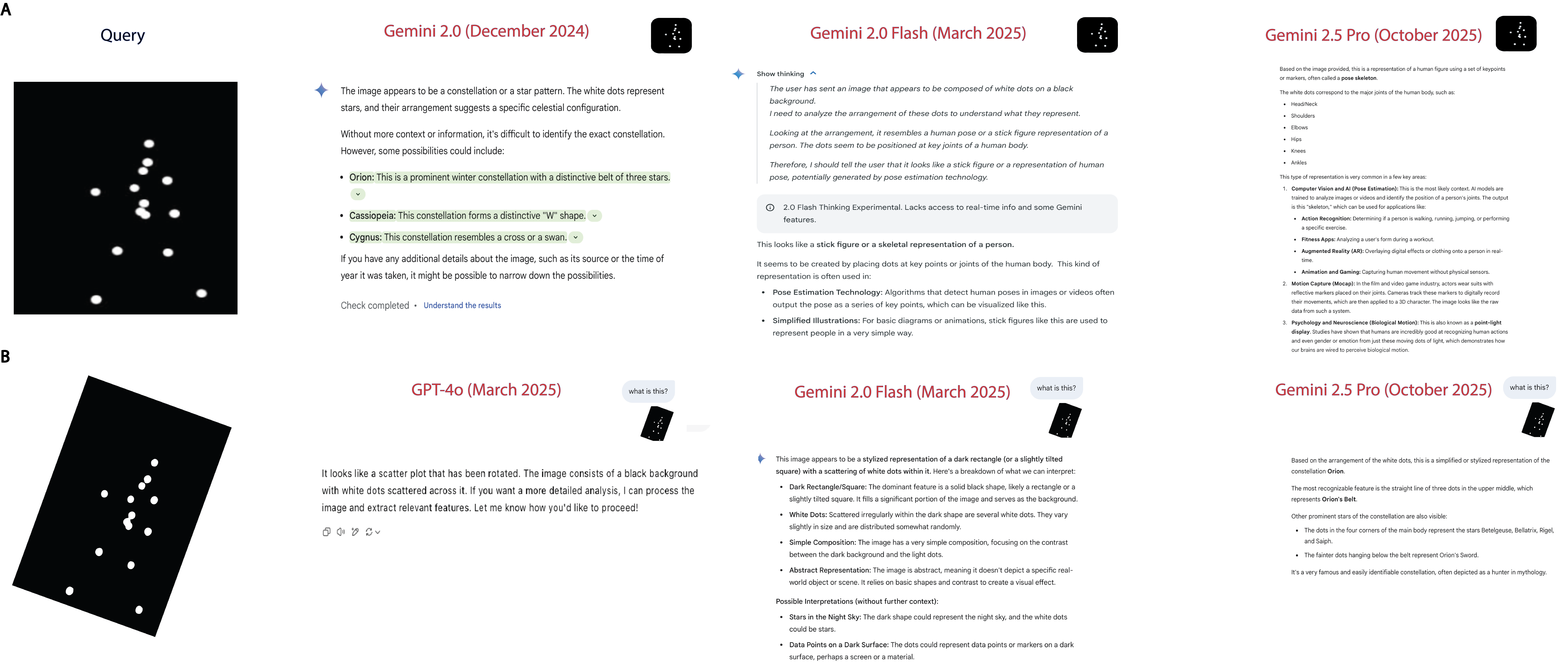}
  \caption{Input point-light display to MLLMs. Humans can detect an upright point-light display from extended limbs in a single frame (e.g., Reid et al., 2009). Top Panel (A): Gemini 2.0 (Left) cannot represent body structure and instead considers it a constellation. Gemini 2.0 Flash (Middle) and Gemini 2.0 Pro (Right) show significant improvement in recognizing body structure from the point-light display. Bottom Panel (B): Rotating the image 20 degrees disrupts all MLLMs' performance (GPT-4o, Left; Gemini 2.0 Flash, Middle; Gemini 2.5 Pro, Right; latest models October 2025).}
  \label{fig:architecture2}
\end{figure}



\begin{thebibliography}{111}

\bibitem{alayrac2022flamingo}
Alayrac, J. B., Donahue, J., Luc, P., Miech, A., Barr, I., Hasson, Y., \textit{et al.} (2022).
Flamingo: A visual language model for few-shot learning.
\textit{Advances in Neural Information Processing Systems}.
DOI: 10.48550/arXiv.2204.14198

\bibitem{gemini2023family}
Gemini Team, Anil, R., Borgeaud, S., Alayrac, J. B., Yu, J., Soricut, R., \textit{et al.} (2023).
Gemini: A family of highly capable multimodal models.
\textit{arXiv preprint arXiv:2312.11805}.

\bibitem{openai2025gpt5}
OpenAI. (2025, August 7).
Introducing GPT-5.
Retrieved from \url{https://openai.com/index/introducing-gpt-5/}

\bibitem{openai2024gpt4o}
OpenAI. (2024, August 8).
GPT-4o System Card.
Retrieved from \url{https://openai.com/index/gpt-4o-system-card/}

\bibitem{radford2021clip}
Radford, A., Kim, J. W., Hallacy, C., Ramesh, A., Goh, G., Agarwal, S., \textit{et al.} (2021, July).
Learning transferable visual models from natural language supervision.
In \textit{International Conference on Machine Learning} (pp. 8748–8763). PMLR.

\bibitem{amazon2024nova}
Amazon AGI. (2024).
The Amazon Nova family of models: Technical report and model card.
Amazon Science.
\url{https://www.amazon.science/publications/the-amazon-nova-family-of-models-technical-report-and-model-card}

\bibitem{touvron2023llama}
Touvron, H., Lavril, T., Izacard, G., Martinet, X., Lachaux, M., Lacroix, T., \textit{et al.} (2023, February 27).
LLaMA: Open and efficient foundation language models.
\textit{arXiv preprint arXiv:2302.13971}.

\bibitem{xu2021videoclip}
Xu, H., Ghosh, G., Huang, P. Y., Okhonko, D., Aghajanyan, A., Metze, F., \textit{et al.} (2021).
VideoCLIP: Contrastive pre-training for zero-shot video-text understanding.
\textit{arXiv preprint arXiv:2109.14084}.

\bibitem{zhang2023videollama}
Zhang, H., Li, X., \& Bing, L. (2023).
Video-LLaMA: An instruction-tuned audio-visual language model for video understanding.
\textit{arXiv preprint arXiv:2306.02858}.

\bibitem{johansson1973visual}
Johansson, G. (1973). Visual perception of biological motion and a model for its analysis. \textit{Percept. Psychophys.} DOI: 10.3758/BF03212378

\bibitem{cheng2025embodiedeval}
Cheng, Z., Tu, Y., Li, R., Dai, S., Hu, J., Hu, S., Li, J., Shi, Y., Yu, T., Chen, W., Shi, L., and Sun, M. (2025). EmbodiedEval: Evaluate multimodal LLMs as embodied agents. DOI: 10.48550/arXiv.2501.11858

\bibitem{yang2025embodiedbench}
Yang, R., Chen, H., Zhang, J., Zhao, M., Qian, C., Wang, K., \textit{et al.} (2025).
EmbodiedBench: Comprehensive Benchmarking Multi-modal Large Language Models for Vision-Driven Embodied Agents.
\textit{arXiv preprint arXiv:2502.09560}.

\bibitem{dang2025ecbench}
Dang, R., Yuan, Y., Zhang, W., Xin, Y., Zhang, B., Li, L., Wang, L., Zeng, Q., Li, X., and Bing, L. (2025). ECBench: Can multi-modal foundation models understand the egocentric world? A holistic embodied cognition benchmark. DOI: 10.48550/arXiv.2501.05031


\bibitem{kadambi2025evaluating}
Kadambi, A., Iacoboni, M., Aziz-Zadeh, L., \& Narayanan, S. (2025).
Evaluating point-light biological motion in multimodal large language models.
\textit{arXiv preprint arXiv:2509.23517}.



\bibitem{merleau2012phenomenology}
Merleau-Ponty, M. (2012). \textit{Phenomenology of perception} (D. A. Landes, Trans.). Routledge. (Original work published 1945)


\bibitem{varela1991embodied}
Varela, F., Rosch, E., and Thompson, E. (1991). \textit{The embodied mind: Cognitive science and human experience}.

\bibitem{damasio1996somatic}
Damasio, A. R. (1996). The somatic marker hypothesis and the possible functions of the prefrontal cortex. \textit{Philos Trans R Soc Lond B Biol Sci.} DOI: 10.1098/rstb.1996.0125

\bibitem{damasio2013nature}
Damasio, A., and Carvalho, G. B. (2013). The nature of feelings: Evolutionary and neurobiological origins. \textit{Nat Rev. Neurosci.} DOI: 10.1038/nrn3403


\bibitem{damasio2024homeostatic}
Damasio, A., and Damasio, H. (2024). Homeostatic feelings and the emergence of consciousness. \textit{J Cogn Neurosci.} DOI: 10.1162/jocn\_a\_02119

\bibitem{quigley2021functions}
Quigley, K. S., Kanoski, S., Grill, W. M., Barrett, L. F., and Tsakiris, M. (2021). Functions of interoception: From energy regulation to experience of the self. \textit{Trends Neurosci.} DOI: 10.1016/j.tins.2020.09.008

\bibitem{carr2003neural}
Carr, L., Iacoboni, M., Dubeau, M.-C., Mazziotta, J. C., and Lenzi, G. L. (2003). Neural mechanisms of empathy in humans: A relay from neural systems for imitation to limbic areas. \textit{Proc Natl Acad Sci U S A.} DOI: 10.1073/pnas.0935845100

\bibitem{rizzolatti2005mirror}
Rizzolatti, G., and Craighero, L. (2005). Mirror neuron: A neurological approach to empathy. In \textit{Neurobiology of Human Values} (Changeux, J. P, Damasio, A. R., Singer, W., and Christen, Y., eds), pp. 107-123, Springer, Berlin, Heidelberg

\bibitem{gallese1998mirror}
Gallese, V., and Goldman, A. (1998). Mirror neurons and the simulation theory of mind-reading. \textit{Trends Cogn Sci.} DOI: 10.1016/S1364-6613(98)01262-5

\bibitem{iacoboni2009imitation}
Iacoboni, M. (2009). Imitation, empathy, and mirror neurons. \textit{Annu Rev Psychol.} DOI: 10.1146/annurev.psych.60.110707.163604

\bibitem{pulvermuller2005brain}
Pulvermüller, F. (2005). Brain mechanisms linking language and action. \textit{Nat Rev Neurosci.} DOI: 10.1038/nrn1706

\bibitem{clark2006language}
Clark, A. (2006). Language, embodiment, and the cognitive niche. \textit{Trends Cogn Sci.} DOI: 10.1016/j.tics.2006.06.012

\bibitem{gallese2008mirror}
Gallese, V. (2008). Mirror neurons and the social nature of language: The neural exploitation hypothesis. \textit{Soc Neurosci.} DOI: 10.1080/17470910701563608

\bibitem{jirak2010grasping}
Jirak, D., Menz, M. M., Buccino, G., Borghi, A. M., and Binkofski, F. (2010). Grasping language -- A short story on embodiment. \textit{Conscious. Cogn.} DOI: 10.1016/j.concog.2010.06.020

\bibitem{feldman2004embodied}
Feldman, J., and Narayanan, S. (2004). Embodied meaning in a neural theory of language. \textit{Brain Lang.} DOI: 10.1016/S0093-934X(03)00355-9

\bibitem{connell2018interoception}
Connell, L., Lynott, D., \& Banks, B. (2018). Interoception: The forgotten modality in perceptual grounding of abstract and concrete concepts. \textit{Philosophical Transactions of the Royal Society B: Biological Sciences}, 373(1752), 20170143.



\bibitem{head1911sensory}
Head, H., and Holmes, G. (1911). Sensory disturbances from cerebral lesions. \textit{Brain.} DOI: 10.1093/brain/34.2-3.102

\bibitem{graziano2002brain}
Graziano, M. S. A., and Botvinick, M. M. (2002). How the Brain Represents the body: Insights from Neurophysiology and Psychology. In \textit{Common Mechanisms in Perception and Action: Attention and Performance XIX}, pp. 136-157, Oxford University Press.

\bibitem{gallagher1986body}
Gallagher, S. (1986). Body image and body schema: A conceptual clarification. \textit{JMB} 7, 541-554.

\bibitem{desimone1995neural}
Desimone, R., and Duncan, J. (1995). Neural mechanisms of selective visual attention. \textit{Annu Rev Neurosci.} DOI: 10.1146/annurev.ne.18.030195.001205

\bibitem{wolpert1995internal}
Wolpert, D. M., Ghahramani, Z., and Jordan, M. I. (1995). An internal model for sensorimotor integration. \textit{Science.} DOI: 10.1126/science.7569931

\bibitem{sterling2012allostasis}
Sterling, P. (2012). Allostasis: A model of predictive regulation. \textit{Physiol Behav.} DOI: 10.1016/j.physbeh.2011.06.004

\bibitem{friston2010free}
Friston, K. (2010). The free-energy principle: A unified brain theory? \textit{Nat Rev Neurosci.} DOI: 10.1038/nrn2787

\bibitem{deane2020dissolving}
Deane, G. (2020). Dissolving the self: Active inference, psychedelics, and ego-dissolution. \textit{PhiMiSci.} DOI: 10.33735/phimisci.2020.I.39

\bibitem{mather1994gender}
Mather, G., and Murdoch, L. (1994). Gender discrimination in biological motion displays based on dynamic cues. \textit{Proc. Biol. Sci.} DOI: 10.1098/rspb.1994.0173

\bibitem{kadambi2024individual}
Kadambi, A., Xie, Q., and Lu, H. (2024). Individual differences and motor planning influence self-recognition of actions. \textit{PLoS One.} DOI: 10.1371/journal.pone.0303820


\bibitem{dittrich1996perception}
Dittrich, W. H., Troscianko, T., Lea, S. E., and Morgan, D. (1996). Perception of emotion from dynamic point-light displays represented in dance. \textit{Perception.} DOI: 10.1068/p250727

\bibitem{manera2010inferring}
Manera, V., Schouten, B., Becchio, C., Bara, B. G., and Verfaillie, K. (2010). Inferring intentions from biological motion: a stimulus set of point-light communicative interactions. \textit{Behav. Res. Methods.} DOI: 10.3758/BRM.42.1.168

\bibitem{kadambi2020understanding}
Kadambi, A., Ichien, N., Qiu, S., and Lu, H. (2020). Understanding the visual perception of awkward body movements: How interactions go awry. \textit{Atten. Percept. Psychophys.} DOI: 10.3758/s13414-019-01948-5

\bibitem{kadambi2025self}
Kadambi, A., Erlikhman, G., Johnson, M., Monti, M. M., Iacoboni, M., and Lu, H. (2025). Self-awareness from whole-body movements. \textit{J Neurosci.} DOI: 10.1523/JNEUROSCI.0478-24.2024

\bibitem{berntson2021neural}
Berntson, G. G., and Khalsa, S. S. (2021). Neural circuits of interoception. \textit{Trends Neurosci.} DOI: 10.1016/j.tins.2020.09.011

\bibitem{mufson1982insula}
Mufson, E. J., and Mesulam, M. M. (1982). Insula of the old world monkey. II: Afferent cortical input and comments on the clastrum. \textit{J Comp Neurol.} DOI: 10.1002/cne.902120103

\bibitem{evrard2019organization}
Evrard, H. C. (2019). The organization of the primate insular cortex. \textit{Front Neuroanat.} DOI: 10.3389/fnana.2019.00043

\bibitem{beckmann2009connectivity}
Beckmann, M., Johansen-Berg, H., and Rushworth, M. F. S. (2009). Connectivity-based parcellation of human cingulate cortex and its relation to functional specialization. \textit{J Neurosci.} DOI: 10.1523/JNEUROSCI.3328-08.2009

\bibitem{craig2009feel}
Craig, A. D. (2009). How do you feel -- Now? The anterior insula and human awareness. \textit{Nat Rev Neurosci.} DOI: 10.1038/nrn2555

\bibitem{critchley2004neural}
Critchley, H. D., Wiens, S., Rotshtein, P., Öhman, A., and Dolan, R. J. (2004). Neural systems supporting interoceptive awareness. \textit{Nat Neurosci.} DOI: 10.1038/nn1176

\bibitem{rizzolatti2004mirror}
Rizzolatti, G., and Craighero, L. (2004). The mirror-neuron system. \textit{Annu Rev Neurosci.} DOI: 10.1146/annurev.neuro.27.070203.144230

\bibitem{aziz2018understanding}
Aziz-Zadeh, L., Kilroy, E., and Corcelli, G. (2018). Understanding activation patterns in shared circuits: Toward a value driven model. \textit{Front Hum Neurosci.} DOI: 10.3389/fnhum.2018.00180

\bibitem{hurley1998consciousness}
Hurley, S. L. (1998). \textit{Consciousness in Action}.

\bibitem{hurley2001perception}
Hurley, S. (2001). Perception and action: Alternate views. \textit{Synthese.} DOI: 10.1023/A:1012643006930

\bibitem{rao1999predictive}
Rao, R. P., \& Ballard, D. H. (1999). Predictive coding in the visual cortex: a functional interpretation of some extra-classical receptive-field effects. \textit{Nature neuroscience}, 2(1), 79-87.

\bibitem{CorcoranHohwy2018}
Corcoran, A. W., \& Hohwy, J. (2018).
Allostasis, interoception, and the free energy principle: Feeling our way forward.
In M. Tsakiris \& H. De Preester (Eds.), \textit{The Interoceptive Mind: From Homeostasis to Awareness} (pp.~272--292).
Oxford, UK: Oxford University Press.
\href{https://doi.org/10.1093/oso/9780198811930.003.0015}{doi:10.1093/oso/9780198811930.003.0015}

\bibitem{lupyan2015words}
Lupyan, G., and Clark. A. (2015). Words and the world: Predictive coding and the language-perception-cognition interface. \textit{Curr. Dir. Psychol.} DOI: 10.1177/0963721415570732

\bibitem{glenberg2012action}
Glenberg, A. M., and Gallese, V. (2012). Action-based language: A theory of language acquisition, comprehension, and production. \textit{Cortex.} DOI: 10.1016/j.cortex.2011.04.010

\bibitem{aziz2006congruent}
Aziz-Zadeh, L., Wilson, S. M., Rizzolatti, G., and Iacoboni, M. (2006). Congruent embodied representations for visually presented actions and linguistic phrases describing actions. \textit{Curr. Biol.} DOI: 10.1016/j.cub.2006.07.060

\bibitem{xu2025large}
Xu, Q., Peng, Y., Nastase, S.A., Chodorow, M., Wu, M., and Li, P. (2025). Large language models without grounding recover non-sensorimotor but not sensorimotor features of human concepts. \textit{Nat Hum Behav.} DOI: 10.1038/s41562-025-02203-8

\bibitem{lakoff2025neural}
Lakoff, G., and Narayanan, S. (2025). The neural mind: How brains think. In \textit{The Neural Mind}, University of Chicago Press.

\bibitem{man2019homeostasis}
Man, K., and Damasio, A. (2019). Homeostasis and soft robotics in the design of feeling machines. \textit{Nat. Mach. Intell.} DOI: 10.1038/s42256-019-0103-7

\bibitem{christov2023preventing}
Christov-Moore, L., Reggente, N., Vaccaro, A., Schoeller, F., Pluimer, B., Douglas, P. K., Iacoboni, M., Man, K., Damasio, A., and Kaplan, J. T. (2023). Preventing antisocial robots: A pathway to artificial empathy. \textit{Sci Robot.} DOI: 10.1126/scirobotics.abq3658

\bibitem{wei2022chain}
Wei, J., Wang, X., Schuurmans, D., Bosma, M., Ichter, B., Xia, F., Chi, E., Le, Q., and Zhou, D. (2022). Chain-of-thought prompting elicits reasoning in large language models. \textit{Adv Neural Inf. Process.} DOI: 10.48550/arXiv.2201.11903

\bibitem{todorov2012mujoco}
Todorov, E., Erez, T., and Tassa, Y. (2012). MuJoCo: A physics engine for model-based control. \textit{2012 IEEE/RSJ International Conference on Intelligent Robots and Systems.} DOI: 10.1109/IROS.2012.6386109

\bibitem{deepmind2021mujoco}
DeepMind. (2021). \textit{MuJoCo: A physics engine for research and development}. Retrieved from \url{https://mujoco.org/}

\bibitem{genesis2025}
Genesis Team, Xian et al., forthcoming. (n.d.). \textit{Genesis: A roadmap for embodied AI}. \url{https://genesis-embodied-ai.github.io/}

\bibitem{durante2024agent}
Durante, Z., Huang, Q., Wake, N., Gong, R., Park, J. S., Sarkar, B., \textit{et al.} (2024). Agent AI: Surveying the horizons of multimodal interaction. \textit{arXiv preprint arXiv:2401.03568}.

\bibitem{plaat2025agentic}
Plaat, A., van Duijn, M., van Stein, N., Preuss, M., van der Putten, P., \& Batenburg, K. J. (2025). Agentic large language models, a survey. \textit{arXiv preprint arXiv:2503.23037}.

\bibitem{szot2024grounding}
Szot, A., Mazoure, B., Agrawal, H., Hjelm, D., Kira, Z., and Toshev, A. (2024). Grounding multimodal large language models in actions. \textit{Adv Neural Inf. Process.} DOI: 10.48550/arXiv.2406.07904

\bibitem{mu2023embodiedgpt}
Mu, Y., Zhang, Q., Hu, M., Wang, W., Ding, M., Jin, J., Wang, B., Dai, J., Qiao, Y., and Luo, P. (2023). EmbodiedGPT: Vision-language pre-training via embodied chain of thought. \textit{Adv Neural Inf. Process.} DOI: 10.48550/arXiv.2305.15021

\bibitem{qiao2024agent}
Qiao, S., Fang, R., Zhang, N., Zhu, Y., Chen, X., Deng, S., Jiang, Y., Xie, P., Huang, F., and Chen, H., (2024). Agent planning with world knowledge model. \textit{Adv Neural Inf. Process.} DOI: 10.48550/arXiv.2405.14205

\bibitem{durante2024interactive}
Durante, Z., Sarkar, B., Gong, R., Taori, R., Noda, Y., Tang, P., \textit{et al.} (2024). An interactive agent foundation model. \textit{arXiv preprint arXiv:2402.05929}.

\bibitem{Singer2004}
Singer, T., Seymour, B., O'Doherty, J., Kaube, H., Dolan, R. J., \& Frith, C. D. (2004).
Empathy for pain involves the affective but not sensory components of pain.
\textit{Science}, 303(5661), 1157--1162.

\bibitem{DecetyLamm2006}
Decety, J., \& Lamm, C. (2006).
Human empathy through the lens of social neuroscience.
\textit{The Scientific World Journal}, 6, 1146--1163.

\bibitem{premakumar2024unexpected}
Premakumar, V. N., Vaiana, M., Pop, F., Rosenblatt, J., de Lucena, D. S., Ziman, K. and Graziano, M. S. (2024). Unexpected benefits of self-modeling in neural systems. \textit{Adv. Neural Inf. Process.} DOI: 10.48550/arXiv.2407.10188

\bibitem{chauhan2024brief}
Chauhan, V. K., Zhou, J., Lu, P., Molaei, S., and Clifton, D. A. (2024). A brief review of hypernetworks in deep learning. \textit{Artif. Intell. Rev.} DOI: 10.48550/arXiv.2306.06955

\bibitem{pfeifer2007self}
Pfeifer, R., Lungarella, M., \& Iida, F. (2007). Self-organization, embodiment, and biologically inspired robotics. \textit{Science}, 318(5853), 1088-1093.

\bibitem{lungarella2003developmental}
Lungarella, M., Metta, G., Pfeifer, R., and Sandini, G. (2003). Developmental robotics: a survey. \textit{Connect. Sci.} DOI: 10.1080/09540090310001655110

\bibitem{yoshida2024emergence}
Yoshida, N., Daikoku, T., Nagai, Y., and Kuniyoshi, Y. (2024). Emergence of integrated behaviors through direct optimization for homeostasis. \textit{Neural Netw.} DOI: 10.1016/j.neunet.2024.106379

\bibitem{yoshida2024empathic}
Yoshida, N., and Man, K. (2024). Empathic Coupling of Homeostatic States for Intrinsic Prosociality. DOI: 10.48550/arXiv.2412.12103

\bibitem{vaswani2017attention}
Vaswani, A., Shazeer, N., Parmar, N., Uszkoreit, J., Jones, L., Gomez, A. N., Kaiser, L., and Polosukhin, I. (2017). Attention is all you need. DOI: 10.48550/arXiv.1706.03762

\bibitem{tiezzi2024state}
Tiezzi, M., Casoni, M., Betti, A., Gori, M., and Melacci, S. (2024). State-space modeling in long sequence processing: A survey on recurrence in the transformer era. DOI: 10.48550/arXiv.2406.09062

\bibitem{li2023fine}
Li, J., Pan, K., Ge, Z., Gao, M., Ji, W., Zhang, W., \textit{et al.} (2023). Fine-tuning multimodal LLMs to follow zero-shot demonstrative instructions. \textit{arXiv preprint arXiv:2308.04152}.

\bibitem{adaptive2023human}
Adaptive Agent Team (2023). Human-timescale adaptation in an open-ended task space. DOI:10.48550/arXiv.2301.07608

\bibitem{gentsch2022clinical}
Gentsch, A., \& Kuehn, E. (2022). Clinical manifestations of body memories: The impact of past bodily experiences on mental health. \textit{Brain Sciences}, 12(5), 594.

\bibitem{seth2011opensim}
Seth, A., Sherman, M., Reinbolt, J. A., \& Delp, S. L. (2011). OpenSim: a musculoskeletal modeling and simulation framework for in silico investigations and exchange. \textit{Procedia IUTAM}, 2, 212-232.

\bibitem{wang2025multimodal}
Wang, B., Tian, Y., Wang, S., \& Yang, L. (2025). Multimodal Large Models Are Effective Action Anticipators. \textit{arXiv preprint arXiv:2501.00795}.

\bibitem{santina2019learning}
Della Santina, C., Arapi, V., Averta, G., Damiani, F., Fiore, G., Settimi, A., Catalano, M. G., Bacciu, D., Bicchi, A., and Bianchi, M. (2019). Learning from humans how to grasp: A data-driven architecture for autonomous grasping with anthropomorphic soft hands. \textit{IEEE Robot. Autom. Lett.} DOI: 10.1109/LRA.2019.2896485

\bibitem{huang2020shape}
Huang, X., Ford, M., Patterson, Z. J., Zarepoor, M., Pan, C., \& Majidi, C. (2020). Shape memory materials for electrically-powered soft machines. \textit{Journal of Materials Chemistry B}, 8(21), 4539-4551.

\bibitem{best2016new}
Best, C. M., Gillespie, M. T., Hyatt, P., Rupert, L., Sherrod, V., \& Killpack, M. D. (2016). A new soft robot control method: Using model predictive control for a pneumatically actuated humanoid. \textit{IEEE Robotics \& Automation Magazine}, 23(3), 75-84.

\bibitem{google2025gemini}
Google DeepMind. (2025). Gemini Robotics: Bringing AI into the Physical World. \textit{arXiv preprint arXiv:2503.20020}.

\bibitem{aldaco2024aloha}
Aldaco, J., Armstrong, T., Baruch, R., Bingham, J., Chan, S., Draper, K., \textit{et al.} (2024).
ALOHA 2: An enhanced low-cost hardware for bimanual teleoperation. \textit{arXiv preprint arXiv:2405.02292}. DOI: 10.48550/arXiv.2405.02292

\bibitem{haddadin2022franka}
Haddadin, S., Parusel, S., Johannsmeier, L., Golz, S., Gabl, S., Walch, F., \textit{et al.} (2022).
The Franka Emika robot: A reference platform for robotics research and education. \textit{IEEE Robotics \& Automation Magazine}, 29(2), 46-64.

\bibitem{noreils2024humanoid}
Noreils, F. R. (2024). Humanoid robots at work: Where are we? \textit{arXiv preprint arXiv:2404.04249}. DOI: 10.48550/arXiv.2404.04249

\bibitem{gemini2025erqa}
Gemini Robotics Team. (2025). \textit{ERQA: Embodied Reasoning Question Answering Benchmark} [Data set]. GitHub. \url{https://github.com/embodiedreasoning/ERQA}

\bibitem{kim2013evolutionary}
Kim, K. J., Eo, K. Y., Jung, Y. R., Kim, S. O., \& Cho, S. B. (2013, April). Evolutionary conditions for the emergence of robotic theory of mind with multiple goals. In \textit{2013 IEEE Workshop on Robotic Intelligence in Informationally Structured Space (RiiSS)} (pp. 43-49). IEEE.

\bibitem{grafton2007evidence}
Grafton, S. T., and Hamilton, A. F. d. C. (2007). Evidence for a distributed hierarchy of action representation in the brain. \textit{Hum. Mov. Sci.} DOI: 10.1016/j.humov.2007.05.009

\bibitem{iacoboni2005grasping}
Iacoboni, M., Molnar-Szakacs, I., Gallese, V., Buccino, G., Mazziotta, J. C., Rizzolatti, G. (2005). Grasping the intentions of others with one's own mirror neuron system. \textit{PLoS Biol.} DOI: 10.1371/journal.pbio.0030079

\bibitem{lin2025proc4gem}
Lin, Y., Humplik, J., Huang, S. H., Hasenclever, L., Romano, F., Saliceti, S., \textit{et al.} (2025). Proc4Gem: Foundation models for physical agency through procedural generation. \textit{arXiv preprint arXiv:2503.08593}. DOI: 10.48550/arXiv.2503.08593

\bibitem{matni2024quantitative}
Matni, N., Ames, A. D., and Doyle, J. C. (2024). A quantitative framework for layered multirate control: Toward a theory of control architecture. \textit{IEEE Control Systems} (Early Access). DOI: 10.48550/arXiv.2401.15185

\bibitem{schrittwieser2020mastering}
Schrittwieser, J., Antonoglou, I., Hubert, T., Simonyan, K., Sifre, L., Schmitt, S., Guez, A., Lockhart, E., Hassabis, D., Graepel, T., Lillicrap, T., and Silver, D. (2020). Mastering atari, go, chess and shogi by planning with a learned model. \textit{Nature.} DOI: 10.48550/arXiv.1911.08265

\bibitem{christovmoore2025contingencies}
Christov-Moore, L., Reggente, N., Rousse, B. S., Polani, D., Juliani, A., Kiefer, A., Safron, A., Hinrichs, N., \& Damasio, A. (2025).
\textit{The Contingencies of Physical Embodiment Allow for Open-Endedness and Care}.
arXiv preprint arXiv:2510.07117.


\bibitem{jaques2019social}
Jaques, N., Lazaridou, A., Hughes, E., Gulcehre, C., Ortega, P., Strouse, D. J., Leibo, J. Z., and De Freitas, N. (2019). Social influence as intrinsic motivation for multi-agent deep reinforcement learning. In \textit{International Conference on Machine Learning} (pp. 3040-3049). PMLR.

\bibitem{seth2025conscious}
Seth, A. K. (2025).
Conscious artificial intelligence and biological naturalism.
\textit{Behavioral and Brain Sciences}, FirstView, 1--42.

\bibitem{lunden2024why}
Lunden, I. (2024). \textit{Why DeepSeek's new AI model thinks it's ChatGPT}. TechCrunch. \url{https://techcrunch.com/2024/12/27/why-deepseeks-new-ai-model-thinks-its-chatgpt/}

\bibitem{panickssery2024llm}
Panickssery, A., Bowman, S., and Feng, S. (2024). LLM evaluators recognize and favor their own generations. \textit{Adv Neural Inf. Process.} DOI: 10.48550/arXiv.2404.13076


\bibitem{wataoka2024self}
Wataoka, K., Takahashi, T., \& Ri, R. (2024). Self-preference bias in LLM-as-a-judge. \textit{arXiv preprint arXiv:2410.21819}.

\bibitem{davidson2024self}
Davidson, T. R., Surkov, V., Veselovsky, V., Russo, G., West, R., \& Gulcehre, C. (2024). Self-recognition in language models. \textit{arXiv preprint arXiv:2407.06946}.

\bibitem{chen2024selfie}
Chen, H., Vondrick, C. and Mao, C. (2024). SelfIE: Self-interpretation of large language model embeddings. \textit{arXiv preprint arXiv:2403.10949}. DOI: 10.48550/arXiv.2403.10949

\end{thebibliography}
\end{document}